\documentclass[aps]{revtex4}
\usepackage{amsfonts}
\usepackage{amsmath}
\usepackage{amssymb}
\usepackage{graphicx}

\input{tcilatex}

\begin{document}

\title[Selective heating]{Selective heating of the ferroelectric film soft
mode phonons}
\author{A.V. Prudan}
\affiliation{Electrotechnical University, Prof. Popov str. 5, 197376 , St. Petersburg,
Russia}
\author{A.V. Mezenov}
\affiliation{Electrotechnical University, Prof. Popov str. 5, 197376, St. Petersburg,
Russia}
\author{S.A. Ktitorov}
\affiliation{A.F. Ioffe Physicotechnical Institute, Polytechnicheskaja str. 26, 194021
St. Petersburg, Russia; Electrotechnical University, Prof. Popov str. 5,
197376, St. Petersburg, Russia}
\keywords{soft mode phonons, two-liquid hydrodynamics, pumping}

\begin{abstract}
Results of the experimental study of the electromagnetic pumping of
frequency 0.3 THz upon the soft mode phonons in the (Ba, Sr)TiO$_{3}$ film
are presented. Some features of the phonon state are revealed using the
capacitor thermometer and the thermocouple. The soft mode phonon overheating
estimated comparing changes of the planar capacitor capacitance was observed
to be quite significant
\end{abstract}

\maketitle










\section{Introduction}

The dielectric susceptibility $\epsilon$ of a ferroelectric is determined by
the phonon spectrum of the crystal and by the phonon state distribution
functions (outside the harmonic approximation) \cite{Vaks}. This leads to a
temperature dependence of $\epsilon$ in the equilibrium case. However, there
exists in principle a possibility to create an nonequilibrium phonon
distribution that can allow us to control the crystal properties with small
delay and applying rather weak controlling fields. In particular, a
nonequilibrium state can be created with the high-frequency electromagnetic
radiation, which interacts selectively with the soft mode phonons. As a
precedent we can refer to the discussion of possible stimulation of the
superconductor transition by the electromagnetic radiation \cite{Gur}.

A study of the perovskite ferroelectrics with the IR and Raman spectroscopy
methods gives information on the frequency and attenuation of the soft mode
and reveals dependence of these parameters on the structure defects and the
chemical content fluctuations \cite{petzelt1}, \cite{petzelt2}, \cite%
{petzelt3}, \cite{Tenne}. Characteristic frequency of the dielectric
susceptibility (and, therefore, of the soft mode) in the SrTiO$_{3}$ films
measured at the room temperature lies within the range of (0,4 $\div$\ 0,7)
THz). The dispersion frequencies observed in the Ba$_{x}$Sr$_{1-x}$TiO$_{3}$
solid solution films lie in the same range \cite{Tkach}. The relaxation time 
$\tau$ and the inverse soft mode frequency $\omega_{0}^{-1}$ for the
oxygen-octahedral ferroelectrics measured by the IR spectroscopy method are
of the same order: $\omega_{0}\tau\thicksim1$. This product is small ($%
\omega _{0}\tau<1$) near the phase transition point: overdamped mode. As the
temperature shifts away from this point, the inequality reverses its sign
that indicates a transition of the soft mode into the underdamped state.
Notice that films have a wider temperature range, where the soft mode is
overdamped in comparison with monocrystals \cite{Tenne}. Theory predicts 
\cite{Vaks} that at approaching the phase transition point in monocrystals $%
\omega_{0}^{2}$ tends to zero as $\left( T-T_{c}\right) $, while $\tau\left(
T\right) ^{-1}\thicksim\frac{aT}{\hbar\omega_{0}}+b\left( \frac{T}{%
\hbar\omega_{0}}\right) ^{2}.$ Here $a$ and $b$ are parameters of the
material \cite{Vaks}.

Results of the study of the soft mode phonons in the ferroelectric thin film
by the electromagnetic pumping of the one-millimeter band are presented in
this paper.

Ba$_{0.4}$Sr$_{0.6}$TiO$_{3}$ solid solution film was deposited on the
dielectric substrate. Within the framework of the harmonic approximation an
ideal interface between the film and the substrate is practically
transparent for the acoustic phonons (neglecting rather weak reflection due
to the sound velocities mismatch) and ideally reflecting for the soft mode
phonons. Taking into account the anharmonicity and violation of the
quasimomentum conservation law in the vicinity of the interface leads to
conclusion that the energy exchange between the phonon subsystems is
anomalously intensive in this domain. The substrate plays a role of the
thermostat due to good thermal contact with the environment. The thermal
capacity of the thermostat $C$ essentially exceeds one of soft mode phonon
subsystem $C_{s}.$ Smallness of $C_{s}$ in comparison with $C$ provides a
necessary condition for overheating of the soft mode subsystem by the
pumping field, while other oscillatory modes remain practically unperturbed.
Preliminary analysis shows that a condition of the soft mode overheating
reads:

\begin{equation}
\frac{C_{f}}{C_{s}}\cdot\frac{\tau_{s}}{\tau_{f}}>1,
\label{thermalcapacities}
\end{equation}
where $C_{s}$ and $C_{f}$ are respectively the critical softmode and
noncritical subsystem thermal capacities; $\tau_{s}$ and $\tau_{f}$ are
respectively the characteristic times for energy exchange between the modes
and the film and the substrate. The integral parameters $C_{f}$ and can be
measured and calculated with good accuracy \ref{kozyrev}. Two other values $%
C_{s}$ and $\tau_{s}$ are very difficult to determine. The uncertainty stems
mainly from unknown phonon-phonon and phonon-impurity interaction constants.
Comprehensive study of microscopic interaction mechanisms is necessary.

\section{Experiment}

A schematic drawing of the experimental setup is depicted in fig. 1.

\bigskip \FRAME{fpFU}{4.9364in}{4.5515in}{0pt}{\Qcb{Experimental setup for
the selective overheating of the soft mode. 1 -- ferroelectric film; 2 --
dielectric substrate; 3 -- electrodes of the planar capacitor; 4 --
thermocouple; k -- wave vector of the electromagnetic pumping}}{\Qlb{fig1}}{%
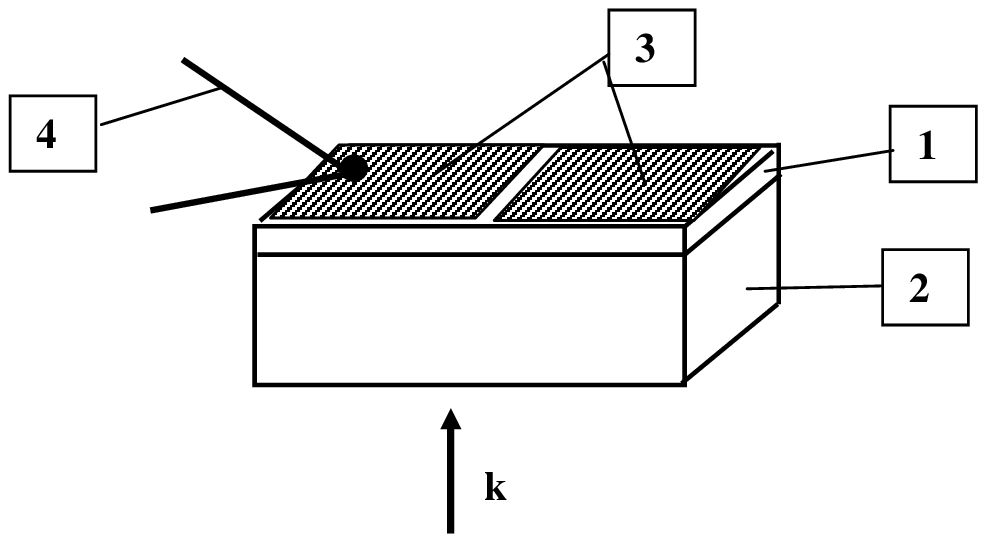}{\special{language "Scientific Word";type
"GRAPHIC";maintain-aspect-ratio TRUE;display "USEDEF";valid_file "F";width
4.9364in;height 4.5515in;depth 0pt;original-width 4.8819in;original-height
4.4987in;cropleft "0";croptop "1";cropright "1";cropbottom "0";filename
'fig1.ps';file-properties "XNPEU";}}

\bigskip

It includes the planar capacitor based on the Ba$_{0.4}$Sr$_{0.6}$TiO$_{3}$
solid solution film 12 $\mu m$ in thickness deposited on the MgO single
crystal substrate 1.5$\times $1.0$\times $0.5 mm in size. Two electrodes are
separated by the narrow gap of 15 $\mu $m. A miniature chromel-copel
thermocouple having one of the junctions in good thermal contact with the
capacitor electrode, was used for measuring the multilayered structure
temperature. The temperature of the soft mode subsystem was monitored
measuring the capacity and using the known capacity dependence on the
totally equilibrium temperature.

Experimental data on the temperature dependence of the inverse capacity of
the samples under the study can be approximated with a good accuracy by the
linear function $C^{-1}\propto\left( T-T_{C}\right) $ in the temperature
range $270\div340K$ \ with $T_{C}=210K.$ Measurements carried out with a use
of the RLC-meter E7-12 provided better than 8\% (for $\Delta T>2K)$
precision of measurements of the temperature increase induced by the pumping
of the critical phonon subsystem. The source of the electromagnetic
radiation created the energy flux in the direction from the substrate
external surface to the ferroelectric film (fig. 1). The MgO slab of the
volume 1.5$\times$1.0$\times $0.5 mm$^{3}$ had a rich spectrum of resonance
frequencies for the electromagnetic field of the millimeter band. Results of
measurement of frequency dependence of the transmittance coefficient $K$
(fig. 2 a) and the pumping induced overheating $\Delta T$ \ in the cases of
the MgO slab with and without the ferroelectric film are presented in fig. 2.

\bigskip

\FRAME{fbpFU}{6.634in}{5.8124in}{0pt}{\Qcb{Dependence of the transmission
coefficient K(f) and the system temperature without (1) and with the
ferroelectric film (2) on the pumping radiation frequency. MgO substrate
(1.5x1.0x0,5) mm3 ; ferroelectric Ba0.4, Sr0.6)TiO3; environment temperature
T = 293 K.}}{\Qlb{fig2}}{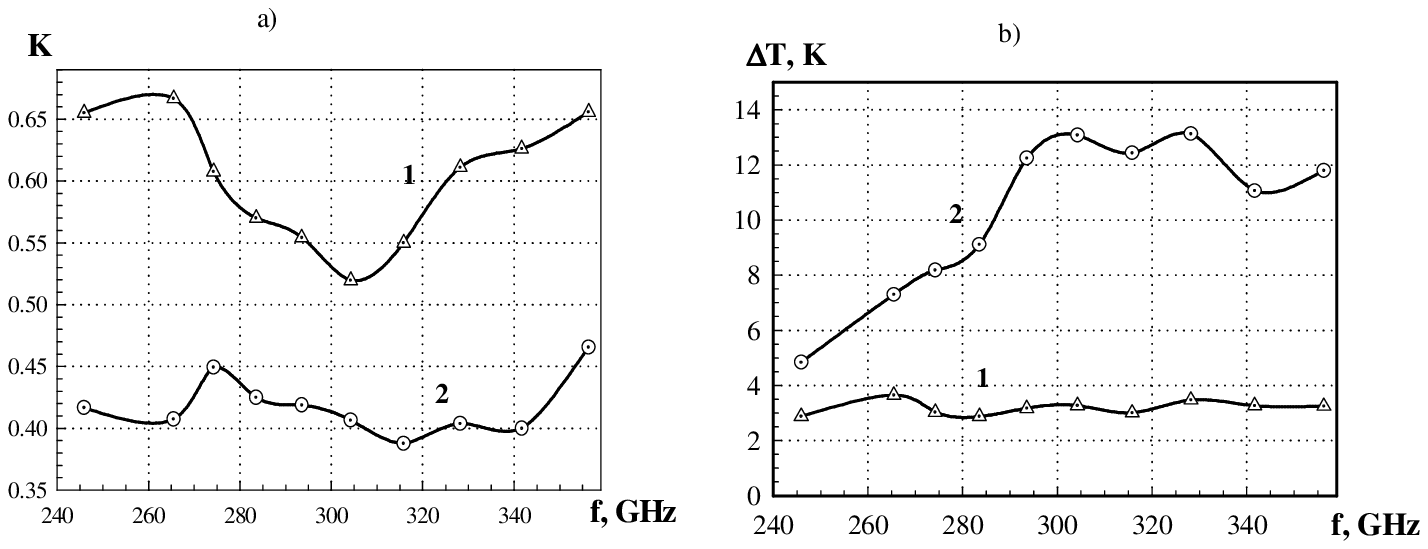}{\special{language "Scientific Word";type
"GRAPHIC";maintain-aspect-ratio TRUE;display "USEDEF";valid_file "F";width
6.634in;height 5.8124in;depth 0pt;original-width 6.6072in;original-height
5.7847in;cropleft "0";croptop "1";cropright "0.9998";cropbottom "0";filename
'fig2.ps';file-properties "XNPEU";}}

Measurements of the equilibrium temperature increase were carried with a use
of the thermocouple at the condition of a weak thermal contact of the sample
with the thermostat. Notice that the frequency dependence $\Delta T(\omega)$
characterizes a spectral behaviour of the integral absorption coefficient of
the whole object under the study.

Comparison of the data presented in fig. 2 b shows that (i) presence of the
ferroelectric film leads to an essential increase of the system
absorptivity; (ii) the overheating $\Delta T(\omega)$ (see fig. 2 b) and,
therefore, the ferroelectric film absorption coefficient depend on the
pumping radiation frequency. The measured frequency dependence of the
overheating $\Delta T(\omega)$ (fig. 2 b) characterizes an integral effect
of the frequency-dependent absorption of the pumping energy by the soft mode
at the condition of the system low Q-factor resonance. Electromagnetic
pumping with the intensity about 6 mW/mm$^{2}$ induces a decrease of the
system capacity. The capacity change magnitude depends on the pumping
frequency and has a maximum, where more, than 20\% decrease of the capacity
occures. Notice that the completely equilibrium overheating of the capacitor
could induce only 10\% decrease of the capacitance. Making the consequent
analysis easier, we presented in fig. 3 the experimental data as an increase
of the structure temperature measured simultaneously with the thermocouple $%
\Delta T$ and the capacitor thermometer $\Delta T_{s}.$

\bigskip \FRAME{fpFU}{4.6553in}{3.9686in}{0pt}{\Qcb{The capacitor
temperature chage\ due to the electromagnetic pumping as a function of the
frequency. (1): capacitor thermpmeter data; (2): thermocouple thermometer
data.}}{\Qlb{fig3}}{fig3.jpg}{\special{language "Scientific Word";type
"GRAPHIC";maintain-aspect-ratio TRUE;display "USEDEF";valid_file "F";width
4.6553in;height 3.9686in;depth 0pt;original-width 12.0832in;original-height
10.2913in;cropleft "0";croptop "1";cropright "1";cropbottom "0";filename
'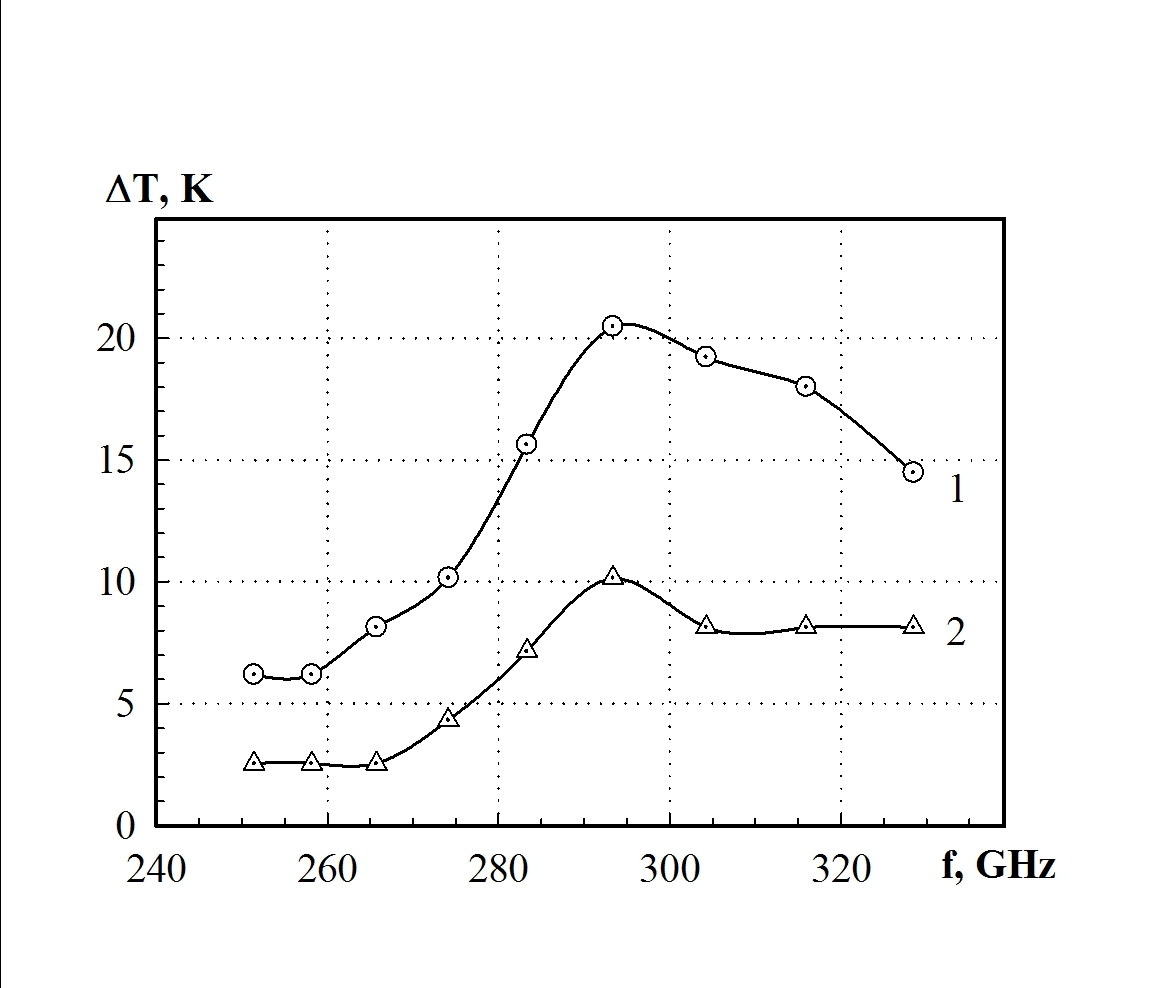';file-properties "XNPEU";}}

\bigskip

Both of the thermometers detected a nonmonotone pumping frequency dependence
of the temperature and demonstrated an essential difference of their data: $%
\Delta T_{s}>\Delta T$ . Maximum temperature difference reached 10 degrees.
Such significant temperature difference cannot be traced back to the
temperature dfference between the thin film absorbing the pumping energy and
the substrate. Numerical consideration of the locally equilibrium
temperature in such a structure predicts significantly lower temperature
difference between the film and the substrate.

\section{Discussion}

This contradiction can be resolved assuming that the pumping induces an
increase of the soft mode phonon occupation numbers exceeding the
equilibrium values. These occupation numbers can be expressed in terms of
the soft mode partial temperature $T$. As it follows from eq. (\ref%
{thermalcapacities}), overheating of the softmode in the ferroelectric is
possible if $C_{s}/C_{f}<\tau _{s}/\tau _{f}.$ This inequality must be made
stronger taking into account magnitudes of the pumping power and the
temperature changes. We believe that the heat capacitance of the soft mode
subsystem can be anomalously small in comparison with one for the entire
system. The relative number of the excited oscillator states near the soft
mode band bottom reads

\begin{subequations}
\begin{equation}
\frac{\delta N}{N}=\left( \frac{\omega _{m}-\omega _{0}}{\omega _{0}}\right)
^{3/2}\left( \frac{\omega _{0}}{\omega _{l}}\right) ^{3},
\label{numberstates}
\end{equation}%
where $\omega _{m}$ is the pumping frequency; $\omega _{0}$ is the soft mode
band lower frequency at the fixed macroscopic state; $\omega _{l}$ is
longitudinal phonon frequency (Debye frequency).

It was assumed that the leading contribution to this temperature dependence
came from the soft mode phonon distribution and, therefore, the capacity
dependence on the soft mode partial temperature in the case of the partial
equilibrium was qualitative similar. Generation of the excess nonequilibrium
soft mode phonons leads to a shift of the soft mode eigenfrequencies $\omega
_{t}(\mathbf{k})$\ in the vicinity of the spectrum bottom and, therefore, to
a change of the dielectric susceptibility: $\epsilon _{0}^{-1}(k=0,\omega
=0)\thicksim \omega _{t0}^{2}(0)+\delta \omega _{t}^{2}(0).$ A deviation of
the squared frequency $\delta \omega _{t}^{2}(0)$ can be estimated using the
self-consistent field approximation: 
\end{subequations}
\begin{equation}
\delta \omega _{t}^{2}(\mathbf{q})=\sum_{j,k}V_{ijqk}\frac{\delta n_{jk}}{%
\omega _{j}\left( \mathbf{k}\right) },  \label{freqshift}
\end{equation}%
where $\delta n_{jk}$ is the excess population number of the nonequilibrium
phonons of the spectrum branch $j$ and the wave vector $\mathbf{k}$, $%
V_{ijqk}$ is the quartic anharmonic interaction constant. The pumping
electromagnetic field and the anharmonic phonon-phonon interaction establish
nonequilibrium state of the critical phonon subsystem belonging to a rather
narrow spectral domain near the dispersion curve $\omega \left( \mathbf{q}%
\right) $ bottom $q=0.$ At high temperatures $\hbar \omega \left( 0\right)
\lesssim k_{B}T$ the nonequilibrium component $\delta n$ of the phonon
distribution function can be written in the form

\begin{equation}
\delta n_{q}=\left\{ 
\begin{array}{rl}
\frac{k_{B}\left( T-T_{0}\right) }{\hbar\omega\left( \mathbf{q}\right) }, & 
\mbox{if $\omega
\left(\mathbf{q}\right)<\omega_{m}$} \\ 
&  \\ 
0, & \mbox{if $\omega\left(\mathbf{q}\right)
>\omega_{m}$}%
\end{array}
\right.  \label{nonequilibrium}
\end{equation}

where $T$ is the partial temperature of the nonequilibrium phonon subsystem; 
$T_{0}$ is the temperature of thermostat formed by noncritical phonons; $%
\hbar \omega _{m}$ is the cut-off energy separating the excited subsystem.
If the pumping frequency $\omega $ exceeds the soft mode lower frequency $%
\omega _{t}\left( 0\right) ,$ then nonequilibrium phonons are located mainly
between these frequencies. The expression (\ref{nonequilibrium}) shows that
the temperature like parameter $T$ can be used as a measure of deviation of
the soft mode phonon subsystem from the equilibrium state. However, the
parameter $T$ cannot be understood too literally as a temperature: it is
rather a parameter characterizing the excess population of the excited
phonon states. Firstly, the local equilibrium conditions, necessary for the
quasihydrodynamical approximation to be valid, are not satisfied; secondly,
the tail of the distribution is not determined by the parameter $T.$ The
overheating $T-T_{0}$ is proportional to a difference of the
susceptibilities $T-T_{0}\propto \epsilon ^{-1}-\epsilon _{0}^{-1}$, which
can be easily measured. The partial temperature can be determined from the
energy balance equation%
\begin{equation}
P=C_{s}\frac{T-T_{0}}{\tau _{e}},  \label{balance}
\end{equation}%
where $\tau _{e}$ is the characteristic time of the energy exchange between
the hot phonons and the thermostat; $C_{s}$ is the partial thermal capacity
of the excited subsystem. The energy balance equation (\ref{balance}) has
significantly wider domain of applicability, than the quasi-hydrodynamic
approximation equations including the energy transport equations.

\section{Conclusion}

In conclusion, the effect of the electromagnetic pumping on the
low-frequency dielectric susceptibility of the thin ferroelectric film is
experimentally observed. A theoretical description of this phenomenon in
terms of the soft mode phonon overheating has been presented. 

\section{Acknowledgements}

The authors acknowledge A. Kozyrev and participants of his seminar for
stimulating discussions. 

This work is supported by the target program "Development of the Higher
School Scientific Potential",  Project RNP 2.12.7083, 2006-2008 years.

\bigskip

\end{document}